\documentclass[prl,twocolumn,groupedaddress]{revtex4}
\usepackage{graphicx,color}
\usepackage{amssymb}   
\usepackage{amsmath}
\usepackage{epstopdf}
\usepackage{natbib}
\usepackage{hyperref}
\usepackage{bm}

\begin{document}
\title{Ising Superconductivity and Majorana Fermions in Transition Metal Dichalcogenides}
\author{Benjamin T. Zhou,  Noah F.Q. Yuan, Hong-Liang Jiang}
\author{K. T. Law} \thanks{phlaw@ust.hk}

\affiliation{Department of Physics, Hong Kong University of Science and Technology, Clear Water Bay, Hong Kong, China }

\begin{abstract}
In monolayer transition metal dichalcogenides (TMDs), electrons in opposite $K$ valleys are subject to opposite effective Zeeman fields, which are referred to as Ising spin-orbit coupling (SOC) fields. The Ising SOC, originated from in-plane mirror symmetry breaking, pins the electron spins to the out-of-plane directions, and results in the newly discovered Ising superconducting states with strongly enhanced upper critical fields. In this work, we show that the Ising SOC generates equal-spin triplet Cooper pairs with spin polarized in the in-plane directions.  Importantly, the spin-triplet Cooper pairs can induce superconducting pairings in a half-metal wire placed on top of the TMD and result in a topological superconductor with Majorana end states. Direct ways to detect equal-spin triplet Cooper pairs and the differences between Ising superconductors and Rashba superconductors are discussed.
\end{abstract} 


\maketitle

\section{\bf I. Introduction}
Monolayer transition metal dichalcogenides (TMDs) are two-dimensional materials composed of a layer of triangularly arranged transition metal atoms sandwiched between two layers of triangularly arranged chalcogenide atoms, forming a 2D honeycomb lattice similar to graphene but with broken sublattice symmetry \cite{Bromley, Boker}. With their strong mechanical properties, relatively high electron mobility, and the massive Dirac energy spectrum \cite{Mak1, Splendiani}, monolayer TMDs are considered potential materials for next generation transistors \cite{Kis, YZhang, QHWang, Bao,Lee}. Interestingly, due to the breaking of in-plane mirror symmetry and the strong atomic spin-orbit coupling (SOC), electrons near the $K$ and $-K$ valleys are subject to strong effective Zeeman fields \cite{Zhu, Xiao, Kormanyos, Zahid, Cappelluti} as depicted in Fig.1a. These effective Zeeman fields strongly polarize electron spins to the out-of-plane directions and the spin polarizations are opposite at opposite valleys. To distinguish this special type of SOC from 2D Rashba SOC which pins electron spins to in-plane directions, we refer to this effective Zeeman field as Ising SOC field. 

Even though the normal states of monolayer TMDs have been studied extensively in recent years \cite{Yao}, the experimental \cite{Ye1, Taniguchi, Shi, Ye2, Saito, Mak4, Mak5}  and theoretical \cite{Ge, Rosner, Roldan, Noah}  studies of the superconducting monolayer TMDs have only just started. It was first shown recently that gated MoS$_2$ thin films, with conducting electrons trapped in a single layer, exhibit superconductivity at about 10K with optimal gating \cite{Ye1,Taniguchi}. Importantly, the in-plane upper critical field $H_{c2}$ of the system can be several times larger than the Pauli limit and an order of magnitude larger than the $H_{c2}$ of bulk superconducting samples where inversion symmetry is restored \cite{Ye2, Saito}. In recent experiments, monolayers of NbSe$_2$ have been successfully fabricated, which are superconducting \cite{Mak4} and the in-plane $H_{c2}$ exhibits strong enhancement similar to gated MoS$_2$ \cite{Mak5}. 

\begin{figure}
\includegraphics[width=3.2in]{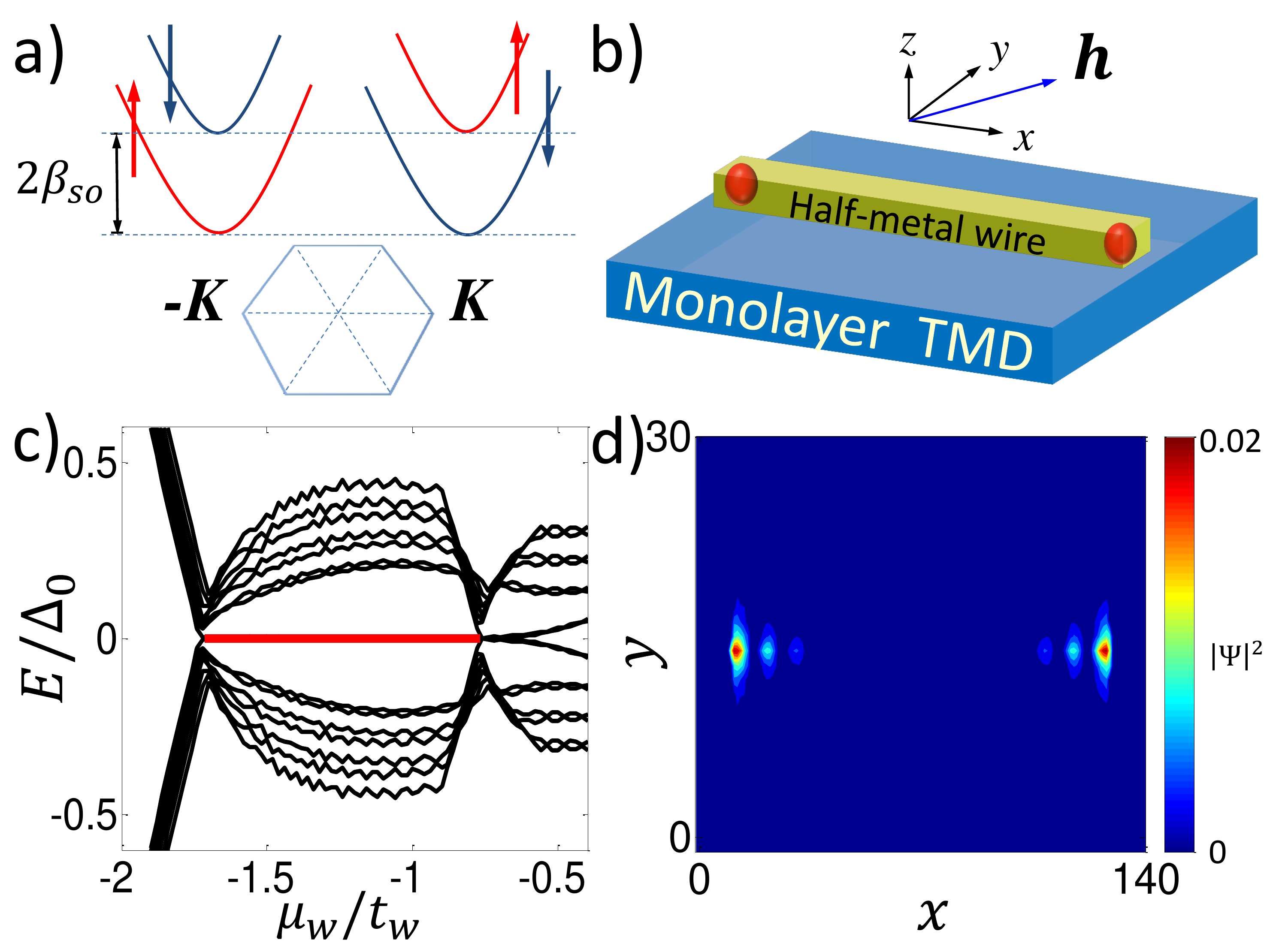}
\caption{a) The electrons near the $K$ and $-K$ valleys subject to the Ising SOC fields in opposite directions. The effective Zeeman gap is $2\beta_{so}$ as defined in Eq.(\ref{h0}). b)  A half-metal wire is placed on top of a superconducting monolayer TMD. Majorana fermions (red dots) appear at the ends of wire when the spin polarization $\bm h$ of the half-metal has in-plane components.  c) The energy spectrum of the set-up in b) as a function of the chemical potential of the wire $ \mu_{w} $, using the tight-binding model in Eq.(8). The red line highlights the topological regime with Majorana modes. d) The zero energy ground state wavefunction of the system in the topological regime in c). Evidently, two Majorana fermions reside at the ends of the wire. The parameters of c) and d) are given in Appendix B.}
\end{figure}

As explained in Refs.\cite{Ye2, Saito, Mak5}, the Ising SOC have opposite directions in opposite valleys so that it preserves time-reversal symmetry and is compatible with superconductivity. Due to the strong pinning of electron spins to the out-of-plane directions by Ising SOC, external in-plane magnetic fields are much less effective in aligning electron spins. As a result, $H_{c2}$ is strongly enhanced. We refer to this special type of non-centrosymmetric superconductor as Ising superconductor \cite{Ye2, Mak5}.

In gated MoS$_2$, the experimental data are well explained by solving the self-consistent gap equation which includes the Ising SOC fields of about 100T ( $\approx$ 6 meV in Zeeman energy) \cite{Ye2, Saito}. In NbSe$_2$, the Ising SOC estimated is even larger at about 660T due to the stronger SOC in the hole bands \cite{Mak5}. Both of the estimations extracted from the MoS$_2$ and NbSe$_2$ experiments are consistent with the corresponding Ising SOC found from DFT calculations \cite{Zhu, Kormanyos, Zahid}.

However, there are still important remaining questions: 1) Besides the in-plane $H_{c2}$ measurements, are there other ways to detect Ising superconductivity? 2) Are there any novel experimental consequences of Ising superconductivity? This work is devoted to answering these two questions.

In the following sections, we show that the Ising SOC induces spin-triplet pairing correlations in an s-wave superconductor. Moreover, the spin-triplet Cooper pairs are formed by electrons with equal spins pointing to in-plane directions. As a result, half-metal leads with spin $\bm h$ polarized to the in-plane directions can freely tunnel Cooper pairs into the superconductor. The Andreev reflection tunneling amplitude decreases as $\bm h$ deviates from the in-plane directions. This would give an experimental signature of Ising superconductivity as depicted in Fig.2. More importantly, when a half-metal wire is placed on top of the Ising superconductor, spin-triplet pairing can be induced on the wire given that $\bm h$ has in-plane components. This would result in a 1D topological superconductor, which supports Majorana end states, as depicted in Fig.1. Finally, the differences between Ising SOC and Rashba SOC are discussed.

\section{\bf II. Equal-spin pairing in Ising superconductors}
To be specific, we study the properties of Ising superconductivity in monolayer MoS$_2$ but the conclusion obtained is very general and can be applied to many other superconducting TMD materials with Ising SOC. In the recent experiments, electrons of MoS$_2$ thin films are mostly trapped in the top layer due to heavy liquid gating. The samples exhibit superconductivity when the conduction bands near the $K$ valleys are filled and the $T_c$ is about 10K at optimal gating \cite{Ye1,Taniguchi,Saito,Ye2}. The conduction bands near the $K$ points are predominantly originating from the Mo $4d_{z^2}$ orbitals of the triangularly arranged Mo atoms \cite{Ye1, Cappelluti}. The effective Hamiltonian near the $K$ valleys, in the basis of $(c_{\bm{k}\uparrow},c_{\bm{k}\downarrow})$, can be written as \cite{Noah}:
\begin{equation} \label{h0}
H_{0}(\bm k=\bm p+\epsilon\bm K)=\left(\dfrac{|\textbf{\textit{p}}|^2}{2m}-\mu\right)\sigma_{0}+\epsilon\beta_{so}\sigma_{z}.
\end{equation}
where $\bm K=(4\pi/3 ,0)$ is the momentum of the $K$ point, $\bm p$ denotes the momentum deviated from $K$ or $-K$ points and $\epsilon = \pm $ is the valley index. The out-of-plane direction is chosen as the $ z $-axis. The $\beta_{so}$ term originates from the coupling between the Mo atoms and the S atoms. As a result, this term breaks the in-plane mirror symmetry. It pins the electron spins to the out-of-plane directions and is referred to as Ising SOC here to distinguish it from the Rashba SOC terms which arise due to mirror symmetry breaking in the out-of-plane direction and pin electron spins to in-plane directions. 

The superconducting MoS$_2$ with spin-singlet s-wave pairing potential $\Delta_0$ can be described by the following mean field Hamiltonian in the Nambu basis $(c_{\bm{k}\uparrow},c_{\bm{k}\downarrow},
c_{-\bm{k}\uparrow}^{\dagger},c_{-\bm{k}\downarrow}^{\dagger})$:
\begin{eqnarray} \label{HBdG}
H_{BdG}(\bm k)=\
\begin{pmatrix}
H_{0}(\bm k) & \Delta_0 i\sigma_y\\
-\Delta_0 i\sigma_y & -H_{0}^{\ast}(-\bm k)
\end{pmatrix}.
\end{eqnarray} 
As demonstrated in the seminal works in Refs.\cite{Gorkov, Sigrist}, the pairing symmetry of the Cooper pairs can be found by solving the Gor'kov equations to obtain the pairing correlations. The pairing correlations are defined as:
\begin{eqnarray}
F_{\alpha \beta}(\bm k,E)=-i\int_{0}^{\infty}e^{i(E+i0^{+})t}\langle\lbrace c_{\bm k, \alpha}(t),c_{-\bm k, \beta}(0) \rbrace\rangle dt.
\end{eqnarray} 
Using $H_{BdG}$ and expressing the pairing correlations in the matrix form, we have:
\begin{eqnarray}
{F}(\bm k,E)=\Delta_0[\psi_{s}(\bm k,E)\sigma_{0}+{\bm d}( \bm k ,E)\cdot {\bm \sigma}]i\sigma_y ,
\end{eqnarray}
where $\psi_{s}$ parametrizes the spin-singlet pairing correlation and the $\bm d$-vector parametrizes the spin-triplet pairing. The $\bm d$-vector is parallel to the $z$-direction in the Ising superconductor case with ${\bm d} = (0, 0, d_z)$. Near the $K$ valleys, 
\begin{eqnarray}
\psi_{s}(\bm p+\epsilon\bm K,E)&=&\frac{E_{+}^2-\Delta _0^2-\xi _{\bm p}^2-\beta _{so}^2}{M(\bm p,E_{+})},\\
d_z (\bm p+\epsilon\bm K,E)&=&\frac{2\epsilon\beta _{so}\xi _{\bm p}}{M(\bm p,E_{+})}
\end{eqnarray}
where $ \xi_{\bm p}=|\bm p|^2/2m-\mu $, $M(\bm p,E)=(\Delta _0^2+\xi _{\bm p}^2-E^2)^2+2 \beta _{so}^2 (\Delta _0^2-\xi _{\bm p}^2-E^2)+\beta _{so}^4$ and $E_{+}=E+i0^{+}$.

It is possible to generalize the pairing matrix of the Hamiltonian in Eq.(2) from $\Delta_0 i\sigma_y$ to $[\Delta_0 + \Delta_t \sigma_z] i\sigma_y$. As shown in Ref.\cite{Noah}, the spin-triplet $\Delta_t$-term belongs to the same irreducible representation as the $\Delta_0$-term. Importantly, this $\Delta_t$-term does not change the form of the pairing correlation matrix in Eq.(4) but enhances the triplet pairing correlation in Eq.(6). The effect of $\Delta_t$ on the pairing correlations is discussed in Appendix A.

It is important to note that for small $\beta_{so}$ the triplet pairing correlation $d_{z}$ is linearly proportional to $\beta_{so}$. As expected, the Ising SOC generates the mixing of spin-singlet and spin-triplet pairings \cite{Gorkov, Sigrist} even when $\Delta_t$ is zero \cite{Sigrist2}. In the basis where the spin quantization axis is along the out-of-plane directions, only the $\sigma_x ,\sigma_y$ components of $ F(\bm k, E) $ are non-zero and both the spin-singlet and spin-triplet Cooper pairs are formed by electrons with opposite spins. However, by choosing the new spin quantization axis in the $ xz $-plane, which forms an angle $\theta$ with the $z$-axis, the pairing correlations become:
\begin{eqnarray} \label{rotate}
{F}_{\theta}(\bm k, E) =
\begin{pmatrix}
-d_z \sin\theta & \psi_{s}+d_z\cos\theta\\
-\psi_{s}+d_z \cos\theta & d_z \sin\theta
\end{pmatrix}, 
\end{eqnarray} 
When the spin quantization axis is along the x-direction with $\theta = \pi/2$, all the triplet Cooper pairs are formed by equal-spin electron pairs with spins pointing to in-plane directions. In the following sections, we show that the property of possessing equal-spin triplet Cooper pairs has important experimental consequences in the detection of Ising superconductivity and the creation of Majorana fermions.

\begin{figure}
\includegraphics[width=3.2in]{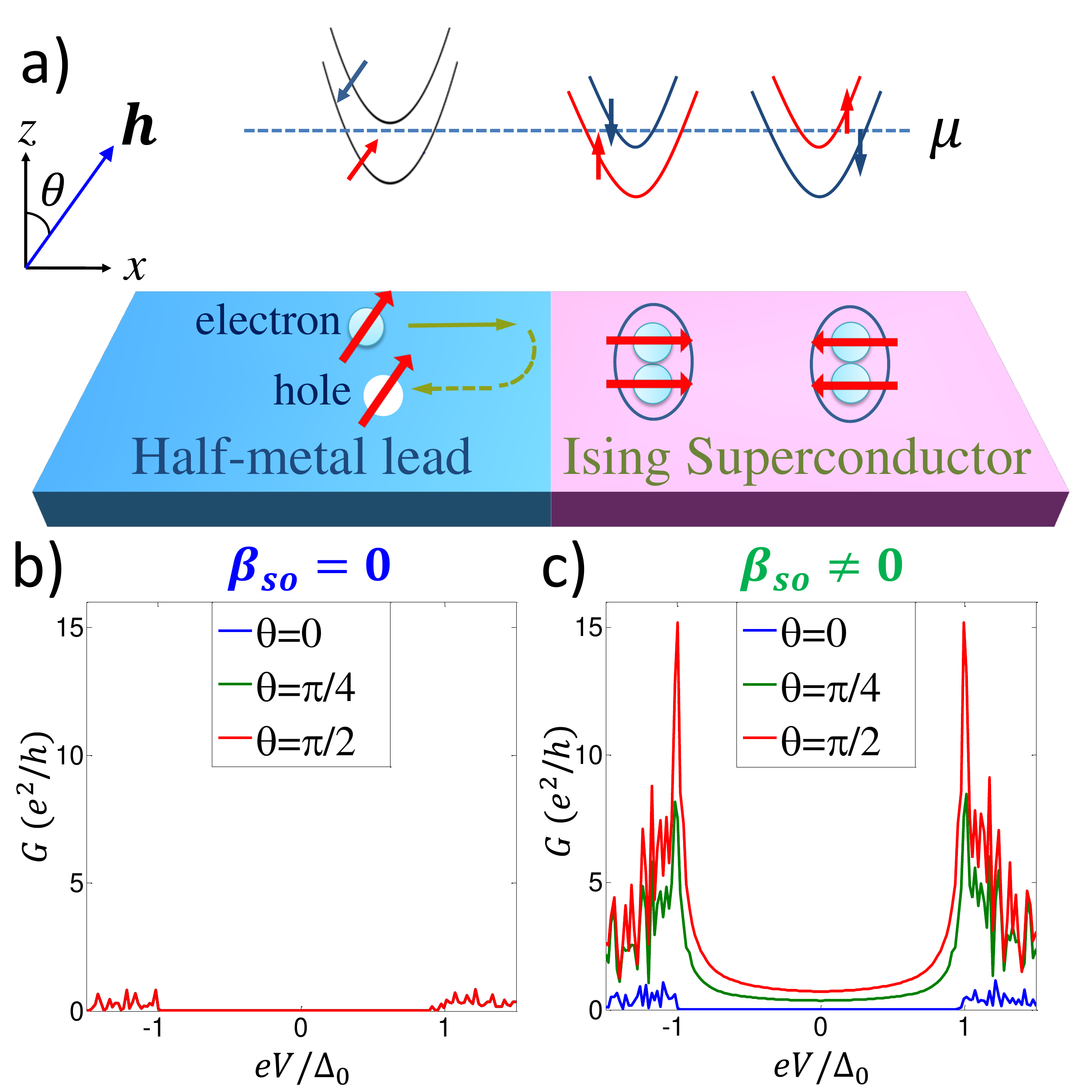}
\caption{a) A half-metal lead is attached to an Ising superconductor. The spin polarization direction of the half-metal is $\bm h$. Both spin-singlet and spin-triplet Cooper pairs can exist in Ising superconductor due to Ising SOC. Equal-spin triplet Cooper pairs have electron spins pointing to in-plane directions. Schematic band structures are shown with the horizontal dashed line representing the chemical potential. b) and c) The tunneling conductance at half-metal/Ising superconductor interface. $\beta_{so} = 0$ in b) and $\beta_{so} \neq 0$ in c). In c), the tunneling conductance decreases when $\bm h$ deviates from the in-plane directions. The parameters of b) and c) are given in Appendix B.}
\end{figure}

\section{\bf III. Detecting equal-spin Cooper pairs} 
As shown in Eq.(\ref{rotate}), the equal-spin triplet pairing correlation is maximum when the spin quantization axis is in-plane and zero when the axis is out-of-plane. When a half-metal lead is attached to the Ising superconductor as depicted in Fig.2a, only equal-spin Andreev reflection processes, which inject equal-spin Cooper pairs into the superconductor, are allowed since all the electrons in the half-metal are spin polarized. On the other hand, ordinary Andreev reflection processes which inject spin-singlet Cooper pairs into the superconductor are strongly suppressed. Since the Cooper pairs in the Ising superconductor have spin pointing to in-plane directions, the equal-spin Andreev reflection amplitude is maximum when the spin polarization $\bm h$ of the half-metal lead is parallel to the in-plane directions. As $\bm h$ deviates from the in-plane directions, the Andreev reflection tunneling amplitude decreases and becomes minimum when $\bm h$ is perpendicular to the in-plane directions.

The tunneling charge conductance of the half-metal/Ising superconductor junction is shown in Fig.2 b)-c), in the absence and presence of the Ising SOC terms respectively. The currents are calculated using recursive Green's function approach \cite{Jie, Nodal} based on the tight-binding model $H_{TMD}$ of Eq.(8) discussed in the next section. From Fig.2b, it is evident that, when $\beta_{so}=0$, the in-gap charge conductance is zero. This is due to the suppression of ordinary Andreev reflection in the half-metal lead. When $\beta_{so}$ is finite, in-gap equal-spin Andreev reflections are possible due to the triplet pairing correlations induced by the Ising SOC and the in-gap conductance is finite. Moreover, the conductance decreases when $\bm h$ deviates from the in-plane direction ($\theta = \pi/2$)  and reaches the minimum when $\bm h$ is perpendicular to the in-plane directions ($\theta = 0 $) as expected. 

If a CrO$2$ film, which has magnetic easy axis pointing to the out-of-plane directions, is used as the half-metal lead, the spin polarization of CrO$2$ can be continuously tuned to the in-plane directions by a small in-plane magnetic field \cite{XiaoG, Goering} which will not induce any orbital effects on the Ising superconductor.

\section{\bf IV. Majorana fermions in Ising superconductor} 
The realization of topological superconductors which support zero energy Majorana bound states has been one of the most important topics in condensed matter physics in recent years \cite{Kane, Qi, Alicea, Beenakker}. One of the most promising ways to realize topological superconductors is to induce pairing by proximity effect on semiconducting wires with Rashba SOC and external magnetic fields \cite{Tewari, Alicea2, ORV, Patrick, Sato, Potter, kou, deng, das}. Magnetic atomic chains on Rashba superconductors can potentially be used to realize Majorana fermions \cite{Yazdani}.

\begin{figure}
\includegraphics[width=3.2in]{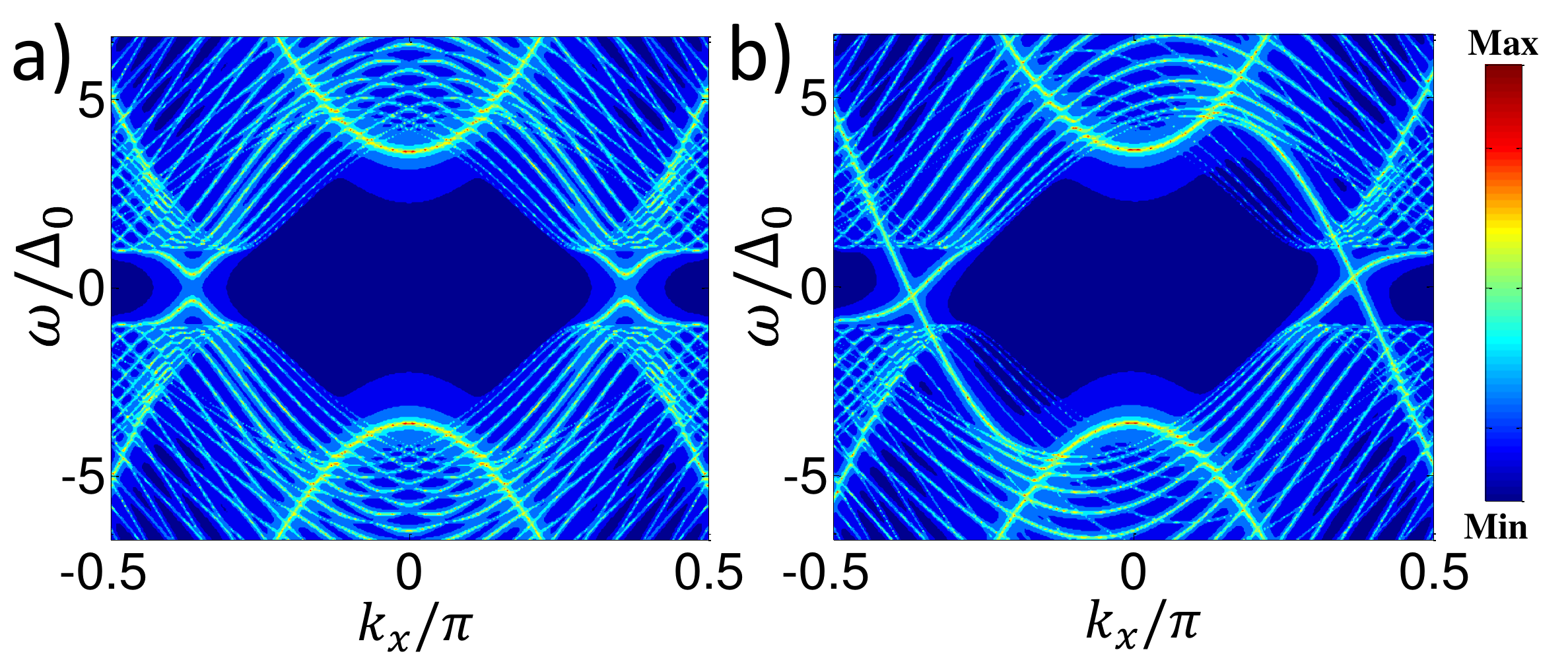}
\caption{ The spectral function of a half-metal wire in proximity to a monolayer superconducting TMD. a) The spin polarization $\bm h$ of the wire is in-plane. An energy gap is induced on the wire. b) $\bm h$ is perpendicular to the plane. The superconducting TMD cannot induce equal-spin pairing on the wire and the wire is gapless. The parameters of the model are given in Appendix B.}
\end{figure}

In this section, we point out that superconducting monolayer TMDs provide an alternative route to realize Majorana fermions. This is due to the fact that in-plane equal-spin triplet pairing correlations are induced by the Ising SOC. When a half-metal wire is placed on top of the TMD, electrons with the same spin can form pairs, as long as the spin polarization of the electrons is not perpendicular to the plane. The induced pairing gap is largest when the spin polarization in the half-metal wire is aligned to in-plane directions.

To demonstrate the induced equal-spin pairing effect on the half-metal wire which is in proximity to the TMD, we consider a system as depicted in Fig.1b. The system is described by the following tight-binding Hamiltonian $H_{tot}$:
\begin{eqnarray} \label{tb}
H_{tot}&=&H_{TMD}+H_{wire}+H_{c}\\\nonumber
H_{TMD}&=&
\sum_{\bm{R},j,s s'} c_{\bm{R},s}^{\dagger}\left(\frac{2}{3m}\sigma_0 +i\frac{\beta_{so}}{3\sqrt{3}}\sigma_{z}\right)_{s s'}c_{\bm{R}+\bm{r}_{j},s'}\\\nonumber
&+&\sum_{\bm{R}}\Delta_{0}c_{\bm{R},\uparrow}^{\dagger}c_{\bm{R},\downarrow}^{\dagger}+h.c.-\sum_{\bm R,s}\left(\mu -\frac{2}{m}\right) c_{\bm{R},s}^{\dagger}c_{\bm{R},s}\\\nonumber
H_{wire}&=&\sum_{n,s}-t_{w}f_{n\bm r_0,s}^{\dagger}f_{(n+1)\bm r_0,s}
-\frac{1}{2}\mu_w f_{n\bm r_0,s}^{\dagger}f_{n\bm r_0,s}\\\nonumber
&+&\sum_{n,s,s'}f_{n\bm r_0,s}^{\dagger}(\frac{1}{2}\bm h\cdot\bm\sigma)_{ss'}f_{n\bm r_0,s'}+h.c.\\\nonumber
H_{c}&=&\sum_{n,s}-t_{c}f_{n\bm r_0,s}^{\dagger}c_{n\bm r_0,s}+h.c.
\end{eqnarray}
where $ c_{\bm R,s} $ and $ f_{\bm R,s} $ are electron annihilation operators of the TMD and the wire respectively. The three lattice vectors of the TMD are denoted as $ \bm r_{j}=(\cos\frac{2j\pi}{3},\sin\frac{2j\pi}{3})\quad (j=0,1,2) $. The hopping and chemical potential of the wire are denoted as $ t_w$ and $\mu_w $ respectively, $ \bm h $ denotes the polarization field in the wire, and $ t_c $ is the coupling between the wire and TMD. 

In order to make the wire a half-metal, we set $ |\mu_w +2t_w | \ll 2|\bm h| $ and the wire is parallel to the zig-zag edge direction of the TMD which is defined as the x-direction. The wire can also be placed along any other directions except the armchair direction where the induced triplet pairing is zero.

Due to the translation symmetry along the wire, we can integrate out the superconducting TMD and plot the spectral function $A(k_{x},\omega)$ of the half-metal wire where
\begin{equation}
A(k_{x},\omega)=\frac{i}{2\pi}\text{tr}[G^{R}(k_{x},\omega)-G^{A}(k_{x},\omega)].
\end{equation}
Here, $G^{R/A}(k_{x}, \omega)$ are the retarded/advanced Green's function of the wire including the self-energy contribution from the superconducting TMD. As shown in Fig.3a, an energy gap opens when the spin polarization $\bm h$ of the wire is parallel to the in-plane direction. On the contrary, the pairing gap vanishes when  $\bm h$ of the half-metal is out-of-plane as shown in Fig.3b. The energy spectrum of the system as a function of $\mu_{w}$, with $\bm h$ pointing to an in-plane direction, is shown in Fig.1c. From Fig.1c, it is evident that there is a topological regime with zero energy modes. The zero energy ground state wavefunction for the system in the topological regime is depicted in Fig.1d. It is evident that there is a Majorana end state residing at each end of the half-metal wire. 

Analytically, after integrating out the TMD background, the effective Hamiltonian $H_{\text{eff}}$ of the half-metal wire at zero frequency can be written as: 
\begin{equation}
H_{\text{eff}}(k_x ,\omega =0)=(-2t_{\text{eff}}\cos k_x -\mu_{\text{eff}})\tau_z +\Delta_{\text{eff}}\sin k_x \tau_x.
\end{equation}
Here, the basis is $ (a_{k_{x}},a_{-k_{x}}^{\dagger}) $ and $ a_{k_{x}} $ is the annihilation operator of a spin polarized electron in the half-metal wire, $ t_{\text{eff}},\mu_{\text{eff}} $ are effective hopping and effective chemical potential respectively, and $ \Delta_{\text{eff}}$ represents the equal-spin pairing induced on the wire by the background TMD. It has the form $ \Delta_{\text{eff}} \propto Z\Delta_{0}\beta_{so} \sin\theta$. This indicates that the Ising SOC term is essential in inducing the pairing on the half-metal wire and $\bm h$ should have in-plane components as shown Fig.3. Importantly, $H_{\text{eff}}$ is the same as the Kitaev model of 1D spinless p-wave superconductor \cite{Kitaev} which is topologically non-trivial when $ |\mu_{\text{eff}} |< 2t_{\text{eff}} $. In other words, the half-metal wire is topological as long as the conduction band is partially occupied. Since the effective Hamiltonian of the system is in the D-class \cite{Schnyder}, introducing small Rashba SOC into the system \cite{Noah,Klinovaja} does not affect the topological phase. It is important to note that $\Delta_{\text{eff}}$ can be comparable with $\Delta_0$ of the parent superconductor. In the particular calculation in Fig.3a, the induced gap is about one-third of $\Delta_0$ which is about 1meV in superconducting NbSe2 or NbS2.

It is important to note that superconducting TMD provides a very practical way to create Majorana fermions. One may place a half-metal wire such as CrO$_2$ on top of a superconducting TMD as depicted in Fig.1b, and then apply an in-plane magnetic field (in any in-plane direction) to align the spins of the half-metal. The half-metal wire can also be replaced by semiconductor wires with large g-factors so that the wire can be easily driven by an in-plane magnetic field to the regime where odd number of transverse subbands are occupied. Another possibility to realize Majorana is to replace the half-metal wire by magnetic atomic chains \cite{Yazdani}. As shown in recent experiments, superconducting TMDs indeed have extremely large in-plane $H_{c2}$, above 50T as shown in Refs. \cite{Ye2, Mak5, Saito}, such that the in-plane magnetic field will not destroy the bulk superconducting properties of the system.

We can now compare Ising superconductors with Rashba superconductors. In 2D Rashba superconductors with Rashba vector $\bm g$ where $\bm g$ is pointing to in-plane directions, the induced equal-spin Cooper pairs in Rashba superconductors have spins aligned in the out-of-plane directions \cite{Sigrist}. When a half-metal wire is placed on top of a Rashba superconductor, superconducting pairing gap can be induced on the wire when the spin polarization of the wire $ \bm h$ has out-of-plane components. This result in a topological superconductor. However, the induced pairing gap vanishes when  $ \bm h$ is in the in-plane direction and perpendicular to the wire. Detailed comparison between Ising and Rashba superconductors can be found in the Appendix C.

Moreover, Ising SOC in TMD materials is very strong and can cause band splitting of over 100meV. We believe that, monolayer or few layers of NbSe$_2$ and NbS$_2$, being intrinsic superconductors with strong Ising SOC in the hole bands \cite{Mak4, Mak5}, are particularly promising materials for realizing Majorana fermions. 

\section{\bf{V. Conclusion}}
In this work, we show that Ising SOC induces equal-spin triplet pairs with electron spins pointing to the in-plane directions. The equal-spin Cooper pairs can be detected in tunneling experiments. Majorana fermions can be created when a half-metal wire or a semiconductor wire is placed on top of a superconducting TMD.

\section{\bf{VI. Ackowledgements}}
KTL thanks Patrick Lee for illuminating discussions. The authors thank the support of HKRGC and Croucher Foundation through HKUST3/CRF/13G, 602813, 605512, 16303014 and Croucher Innovation Grant.

\renewcommand{\theequation}{A-\arabic{equation}}
\setcounter{equation}{0}  
\section*{\textbf{Appendix A: Pairing correlations with both spin-singlet  and spin-triplet order parameters}}

It is well known that inversion symmetry breaking can cause a mixing of spin-singlet and spin-triplet pairing potential\cite{Gorkov, Sigrist, Sigrist2}. In this section, we present the results of the pairing correlation matrix $\hat{F}$ when both singlet and triplet pairing potentials are included in the BdG Hamiltonian in Eq.(\ref{eq:BdG}). We show that the triplet pairing potential will enhance the triplet pairing correlation $d_z$ as defined in Eq.(4) of the main text.

As stated in the main text, by including the usual $s$-wave pairing potential only, the mean-field BdG Hamiltonian for the Ising superconductor in the Nambu basis $\bm{\Psi}=( c_{\bm{k}\uparrow}, c_{\bm{k}\downarrow}, c^{\dagger}_{\bm{-k}\uparrow}, c^{\dagger}_{\bm{-k}\downarrow})^{T}$ takes the form:
\begin{eqnarray}\label{eq:BdG}
H_{BdG}(\bm{k}) = 
\begin{pmatrix}
H_{0}(\bm{k}) & \hat{\Delta}\\
\hat{\Delta}^{\dagger} & -H_{0}^{*}(\bm{-k})
\end{pmatrix}
\end{eqnarray}
where the normal-state Hamiltonian $H_{0}(\bm k=\bm p+\epsilon\bm K)=\xi_{\bm{p}}\sigma_{0}+\epsilon\beta_{so}\sigma_{z}$ as defined in Eq.(1) in the main text, and the pairing matrix $\hat{\Delta}=\Delta_{0}i\sigma_{y}$. From the BdG Hamiltonian in Eq.(\ref{eq:BdG}), one can explicitly show the mixing of spin-singlet and spin-triplet pairing correlations generated by Ising SOC, following Eqs.(3)-(7) in the main text.

As demonstrated in Ref.\cite{Noah}, by taking nearest-neighbor electron-electron attraction into account, it is possible to obtain a finite spin-triplet order parameter $\bm{d_{A1, z}}$ and the pairing matrix in Eq.(\ref{eq:BdG}) is generalized from $\hat{\Delta}=\Delta_{0}i\sigma_{y}$ to $\hat{\Delta}=[ \Delta_{0} + \Delta_{t}\sigma_{z} ] i\sigma_{y}$. The spin-triplet $\Delta_{t}$-term belongs to the same irreducible representation as the spin-singlet $\Delta_{0}$-term. With the spin-triplet order parameter $\bm{d_{A1, z}}$, one can easily generalize the result in Eq.(6) where the triplet-pairing correlation $d_z$ near the two $K$-valleys becomes:
\begin{equation}\label{eq:dz}
d_z (\bm{p},E)=\dfrac{\Delta_{t}P(\bm{p},E_{+}) +2\epsilon \beta_{so} \xi_{\bm{p}}\Delta_0}{Q_{+}(\bm{p},E_{+}) Q_{-}(\bm{p},E_{+})}
\end{equation}  
Here, $P(\bm{p},E)=\Delta_{0}^2-\Delta_{t}^2-\xi_{\bm{p}}^2-4\beta_{so}^2+E^2$, $Q_{+}(\bm{p},E)=(\Delta_0+ \Delta_t)^2+(\xi_{\bm{p}}+\epsilon\beta_{so})^2-E^2$, and $Q_{-}(\bm{p},E)=(\Delta_0-\Delta_t)^2 + (\xi_{\bm{p}}-\epsilon\beta_{so})^2-E^2$, where $\epsilon=\pm$ denotes the valley-index and $E_{+}=E+i0^{+}$. As shown in Eq.(\ref{eq:dz}), the triplet-pairing correlation will only be enhanced by including the triplet-pairing order $\bm{d_{A1, z}}$, and the conclusions in the main text will remain valid.

\renewcommand{\theequation}{B-\arabic{equation}}
\setcounter{equation}{0}  
\section*{\textbf{Appendix B: Tight-binding model parameters}}

In our calculations for Figs.1-3 in the main text, we use the tight-binding Hamiltonian $H_{TMD}$ for the superconducting TMD as shown in Eq.(8) in the main text. The tight-binding parameters for $H_{TMD}$ in Fig.1 c), Fig.2 b)-c) and Fig.3 a)-b) are set as follows: the hopping amplitude $t\equiv -\frac{2}{3m}=-1$ with $m$ denoting the effective mass defined in Eq.(1) in the main text. The Ising SOC splitting $\beta_{so}=0.4t$, the $s$-wave pairing strength $\Delta_{0}=0.15t$, and the chemical potential $\mu=-0.6t$.

In Fig.1 c), the parameters for the half-metal wire are defined in the Hamiltonian $H_F$ in Eq.(8) in the main text and set to be $t_w=-1, |\bm{h}|=0.5$. The coupling between the half-metal wire and the Ising superconductor is $t_c=-1.2t_w$. The Ising superconductor has 140 sites along the zig-zag edge(the edge along the $x$-direction defined in Fig.1 b) and 30 sites along the armchair edge(the edge along the $y$-direction defined in Fig.1 b)). The half-metal wire is 120-site long and is placed parallel to the $x$-direction. In Fig.1 d), we set $\mu_{w}=-1.5t_{w}$ with $\Delta_{0}=0.05t$.

In our calculations for the charge/spin currents in Fig.2 b)-c), we use the following tight-binding Hamiltonian:
\begin{eqnarray} \label{B1}
H_{tot}&=&H_{F}+H_{TMD}+H_{c}\\\nonumber
H_{F}&=&\sum_{\bm{R}',\bm{d},s}-t_{L}(f_{\bm{R}',s}^{\dagger}f_{\bm{R}'+\bm{d},s}+h.c.)\\\nonumber
&-&\sum_{\bm{R}',s}\mu_{L} f_{\bm{R}',s}^{\dagger}f_{\bm{R}',s}+\sum_{\bm{R}',s s'} f_{\bm{R}',s}^{\dagger}(\bm{h}\cdot\bm{\sigma})_{s s'}f_{\bm{R'},s'}\\\nonumber
H_{TMD}&=&\sum_{\bm{R},j,s}-t(c_{\bm{R},s}^{\dagger}c_{\bm{R}+\bm{r}_{j},s} + h.c.)- (\mu+3t) c_{\bm{R},s}^{\dagger}c_{\bm{R},s}\\\nonumber
&+&i\sum_{\bm{R},j,s s'}\frac{\beta_{so}}{3\sqrt{3}} c_{\bm{R},s}^{\dagger}(-1)^{j}(\sigma_{z})_{s s'}c_{\bm{R}+\bm{r}_{j},s'}\\\nonumber
&+&\sum_{\bm{R}}\Delta_{0}(c_{\bm{R},\uparrow}^{\dagger}c_{\bm{R},\downarrow}^{\dagger}+h.c.)\\\nonumber
H_{c}&=&\sum_{\left\langle \bm{R}',\bm{R} \right\rangle, s}-t_{c}(f_{\bm{R}',s}^{\dagger}c_{\bm{R},s}+h.c.)
\end{eqnarray}
where $H_{F}$ represents the Hamiltonian of the half-metal lead and $H_{TMD}$ is defined in the same way for the Ising superconductor as in Eq.(8) in the main text. $H_{c}$ is the coupling Hamiltonian modelling the barrier at the half-metal lead/Ising superconductor interface. The half-metal lead has a square lattice structure. Here, $\bm{R}$ and $ \bm R' $ denote the lattice sites of the Ising superconductor and the lead respectively, $s=\uparrow ,\downarrow$ is the spin index. $ \bm d=\bm x,\bm y $ denotes the primitive vectors of the half-metal lead. The hopping amplitude in the half-metal lead is set to be $t_L=-1$, and $|\bm{h}|=1$. The chemical potential is set at $\mu_L=3.5t_L$ such that all the electron spins are polarized to the same direction at the Fermi energy. The coupling strength at the interface is set to be the same as the hopping amplitude in the lead $t_{c}=t_L$. The Ising superconductor has $60$/$100$ sites along the $x$/$y$ directions respectively. The half-metal lead is attached to the zig-zag edge of the Ising superconductor in our calculations. 

To study the transport properties, we calculate the scattering matrix at the half-metal lead/Ising superconductor interface at energy bias $ E $ using the recursive Green's function method in Refs.\cite{Jie, Nodal}:
\begin{eqnarray}
r_{\alpha\beta}(E) = -I\sigma_0\delta_{\alpha\beta} + i \Gamma^{1/2}_{\alpha} G^{R}(E)\Gamma^{1/2}_{\beta}
\end{eqnarray}
where $\alpha, \beta \in \left\lbrace e,h \right\rbrace$ label the electron or hole, $ \sigma_0 $ is the identity matrix in the spin space, and $ I $ is the identity matrix in the rest of the Hilbert space. $G^{R}(E)=(E+i\eta -H_{tot})^{-1}$ is the retarded Green's function obtained from the tight-binding Hamiltonian in Eq.(\ref{B1}). $\Gamma_{e/h}$ is the electron/hole part of the broadening function. With the scattering matrix,  the charge conductance and the spin conductance in the ${\bm{n}}$ direction at energy $E$ can be readily obtained as:
\begin{eqnarray}
G_{c}(E)&=&\frac{e^2}{h}\text{tr}\lbrace I\sigma_0 - r_{e e}^{\dagger}(E) r_{e e}(E) \\\nonumber
&+& r_{h e}^{\dagger}(E) r_{h e}(E) \rbrace\\\nonumber
G_{s,\pmb{n}}(E)&=&\frac{e^2}{h}\text{tr}\lbrace - r_{e e}^{\dagger}(E)I(\bm{n}\cdot\bm{\sigma} ) r_{e e}(E)\\\nonumber 
&+& r_{h e}^{\dagger}(E)I(\bm{n}\cdot\bm{\sigma}^{*}) r_{h e}(E) \rbrace
\end{eqnarray}
and the total spin conductance is defined as $G_{s, T}= \sqrt{G_{s, x}^2 + G_{s, y}^2 + G_{s, z}^2}$.\\

In Fig.3, the parameters for the half-metal wire is defined in the same way as in Eq.(8) in the main text and set to be $t_w=-1$, $|\bm{h}|=1$, $t_c=-1.2t_w$, and $\mu_w=-1.2t_w$. 

\renewcommand{\theequation}{C-\arabic{equation}}
\setcounter{equation}{0}   
\section*{\textbf{Appendix C: Comparison between Ising superconductors and Rashba superconductors with $s$-wave pairing}}

In this section, we compare Ising superconductors with 2D Rashba superconductors. As studied in Refs.\cite{Gorkov, Sigrist, Sigrist2}, the simplest normal-state Hamiltonian in the presence of Rashba SOC has the generic form:
\begin{equation}
H_{0}=\sum_{\bm{k},ss'}[\xi_{\bm{k}}\sigma_0 +\alpha \bm{g}_{\bm{k}}\cdot \bm{\sigma}]_{ss'}c^{\dagger}_{\bm{k} s}c_{\bm{k} s'}
\end{equation}
with $\xi_{\bm{k}}=\epsilon_{\bm{k}}-\mu$ referring to the kinetic term measured from the chemical potential $\mu$ and the SOC vector $\bm{g}_{\bm{k}}=(k_{y}, -k_{x}, 0)$ . Thus, the Rashba SOC pins the spins of electrons to the in-plane directions, in contrast with the Ising SOC which pins electron spins to the out-of-plane directions. As shown in Refs.\cite{Gorkov, Sigrist, Sigrist2}, assuming an $s$-wave pairing matrix $\hat{\Delta}=\Delta_{0}i\sigma_{y}$, the pairing correlation matrix for Rashba superconductors in the out-of-plane spin basis takes the form: 
\begin{equation}\label{eq:FRashba}
\hat{F}(\bm{k},E_+)=
\begin{pmatrix}
-i\dfrac{k_{-}}{k}\Delta_0 F_{-}(\bm{k},E_+) & \Delta_0 F_{+}(\bm{k},E_+)\\
  -\Delta_0 F_{+}(\bm{k},E_+) & -i\dfrac{k_{+}}{k}\Delta_0 F_{-}(\bm{k},E_+)
\end{pmatrix}
\end{equation}

where $k_{\pm}=k_{x}\pm i k_{y}$, and $F_{\pm}(\bm{k}, E_+)=\frac{1}{2}[(\Delta_{0}^2+(\xi_{\bm{k}}+\alpha |\bm{g_k}|)^2-E_{+}^2)^{-1}\pm (\Delta_{0}^2+(\xi_{\bm{k}}-\alpha |\bm{g_k}|)^2-E_{+}^2)^{-1}]$.

As shown in Eq.(\ref{eq:FRashba}), for Rashba superconductors, the induced equal-spin Cooper pairs have spins pointing to the out-of-plane $z$-direction, with 
\begin{equation}
\begin{split}
F_{\uparrow\uparrow}(\bm{k},E_+)&=-i\Delta_0 F_{-}(\bm{k},E_+)k_{-}/k\\
F_{\downarrow\downarrow}(\bm{k},E_+)&=-i\Delta_0 F_{-}(\bm{k},E_+)k_{+}/k
\end{split}
\end{equation}

\begin{figure}
\includegraphics[width=3.2in]{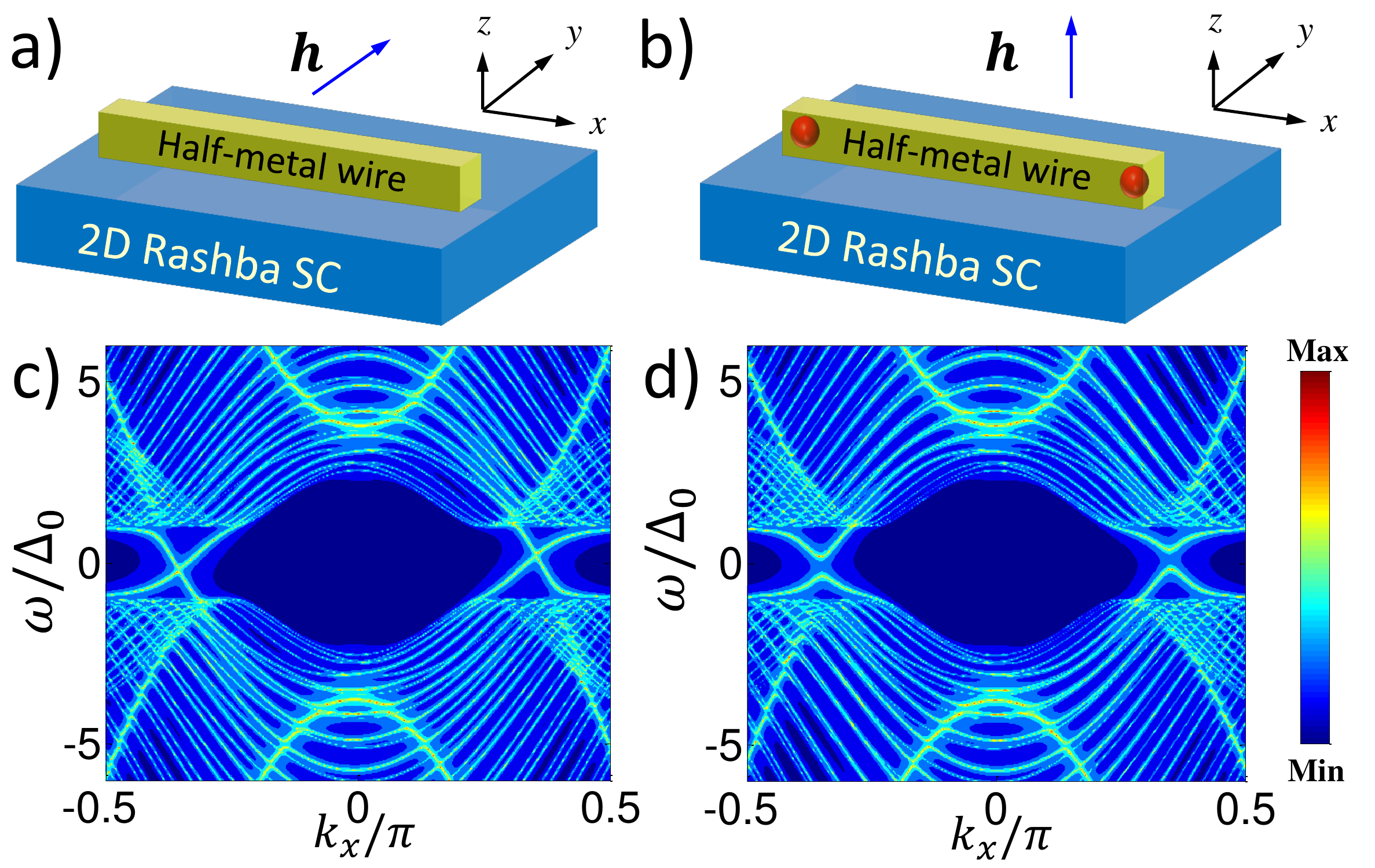}
\caption{The proximity effect of a 2D Rashba superconductor on a half-metal wire. (a) The spin polarization $\bm h$ of the wire is pointing to the in-plane $y$-direction. (b) $\bm h$ is perpendicular to the plane.  (c) The spectral function of the half-metal wire corresponding to the set-up in (a). The Rashba superconductor cannot induce equal-spin pairing on the wire and the wire remains gapless. (d) The spectral function of the half-metal wire corresponding to the set-up in (b). The Rashba superconductor has equal-spin Cooper pairs with spin polarization pointing to the out-of-plane $z$-direction, and thus induces a pairing gap on the half-metal wire.}\label{FIG4}
\end{figure}

From the pairing correlations, one can note that the Rashba superconductor can induce pairings on a half-metal thin film if the spin polarization of the half-metal is pointing to the out-of-plane directions. This will result in a 2D topological superconductor.

To show a specific example of the differences between Ising superconductors and Rashba superconductors, we demonstrate the proximity effect of Rashba superconductors on a half-metal wire as shown in Fig.4. Using the same approach as that in obtaining Fig.3a)-b) in the main text, we calculate the spectral function of the half-metal wire including the self-energy from the 2D Rashba superconductor. The Rashba superconductor is modelled by the following tight-binding Hamiltonian:
\begin{eqnarray}
H_{RSC}&=&\sum_{\bm{R},\bm{r},s}-t(c_{\bm{R},s}^{\dagger}c_{\bm{R}+\bm{r},s} + h.c.)\\\nonumber
&-& (\mu + 3t)c_{\bm{R},s}^{\dagger}c_{\bm{R},s}\\\nonumber
&+& \sum_{\bm{R},\bm{r},s s'}i\alpha_{R} c_{\bm{R},s}^{\dagger}(\bm{r}\times\bm{\sigma})_{s s'}c_{\bm{R}+\bm{r},s'}\\\nonumber
&+&\sum_{\bm{R}}\Delta_{0}(c_{\bm{R},\uparrow}^{\dagger}c_{\bm{R},\downarrow}^{\dagger}+h.c.)
\end{eqnarray}

For comparison, $H_{RSC}$ is the same as $H_{TMD}$ in Eq.\ref{B1} except that the Ising SOC is changed to the Rashba type SOC with $\alpha_{R}=0.2t$. The Hamiltonian of the half-metal wire on the superconductor and the coupling between the Rashba superconductor and the half-metal wire remain unchanged as compared to Eq.\ref{B1} and Eq.8 of the main text. 

 As shown in Fig.4a, when the magnetization direction $\bm{h}$ is in the in-plane direction and perpendicular to the wire, for the Rashba superconductor case, the induced pairing gap on the half-metal wire is zero(Fig.4c). On the contrary, for Ising superconductors, the induced pairing gap on the wire is finite as shown in Fig.3a of the main text. On the other hand, when $\bm{h}$ of the half-metal wire is perpendicular to the plane of the Rashba superconductor, the induced superconducting gap on the half-metal is finite as shown in Fig.4d and thus results in a 1D topological superconductor supporting Majorana end states(red dots in Fig.4b. However, for Ising superconductors, the induced pairing gap is zero in such a case as shown in Fig.3b of the main text.

\end{document}